\documentstyle[aps,epsfig,prl]{revtex}
\begin{document}
\title{ Exact Demonstration of Magnetization Pleateaus 
and First Order Dimer-N\'eel Phase Transitions in a Modified
Shastry-Sutherland Model for SrCu$_2$(BO$_3$)$_2$}
\author{Erwin M\"uller-Hartmann\cite{byline1} and 
Rajiv R. P. Singh \cite{byline2}}
\address{Institute for Theoretical Physics, University of California,
Santa Barbara, CA 93106}
\author{Christian Knetter and G\"otz S. Uhrig}
\address{Institut f\"ur Theoretische Physik, Universit\"at zu K\"oln, 
Z\"ulpicher Str. 77, D-50937 K\"oln, Germany}
\twocolumn[\hsize\textwidth\columnwidth\hsize\csname
@twocolumnfalse\endcsname
\date{\today}
\maketitle
\widetext
\begin{abstract}
\begin{center}
    \parbox{6in} {We study a generalized Shastry-Sutherland 
model for the material SrCu$_2$(BO$_3$)$_2$.
Along a line in the parameter space,
we can show rigorously that the model
has a first order phase transition between
Dimerized and N\'eel-ordered ground states.
Furthermore, when a magnetic field is applied in the Dimerized
phase, magnetization plateaus develop at commensurate values
of the magnetization.
We also discuss various aspects
of the phase diagram and properties of this model away from this
exactly soluble line, which include gap-closing continuous transitions
between Dimerized and magnetically ordered phases.
}
\end{center}
\end{abstract}
\pacs{}

]

\narrowtext

In recent years, many novel magnetic materials have been synthesized
which exhibit spin-gap behavior. In these materials the ground state is a
spin-singlet and there is a gap to all spin excitations. Such phenomena
have long been studied in quasi-one dimensional systems, but much recent
interest has arisen from the discovery of quasi-two dimensional spin-gap
materials CaV$_4$O$_9$ \cite{taniguchi}, 
Na$_2$Ti$_2$Sb$_2$O \cite{natisbo} and SrCu$_2$(BO$_3$)$_2$ \cite{kageyama}. 
The latter material is particularly interesting in that,
by virtue of the crystal geometry, it is an experimental realization
of the Shastry-Sutherland model \cite{shastry}, for which an exact dimerized
singlet eigenstate can be written down, which for a range of parameters
is the ground state of the model. 
Among the interesting experimental findings for SrCu$_2$(BO$_3$)$_2$
are that the system appears very close to a transition to a N\'eel
phase and it also shows magnetization plateaus as a function of
magnetic field \cite{kageyama,miyahara}.

\begin{figure}
\epsfxsize=65mm
\centerline{\epsffile{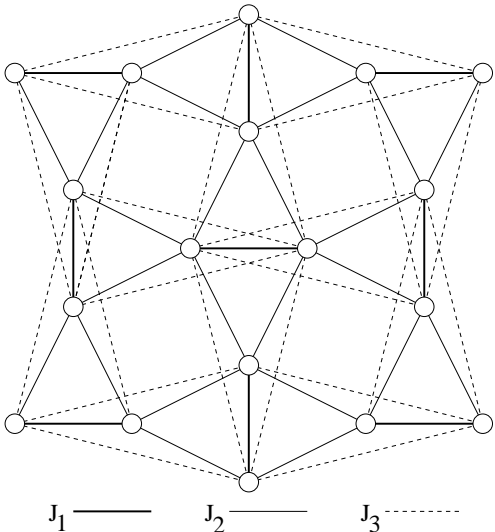}}
\caption[*]{The geometry of spins and
couplings in the material SrCu$_2$(BO$_3$)$_2$.
}
\label{Fig.1}
\end{figure}

Here we consider a generalized Shastry-Sutherland model, with Hamiltonian
\begin{equation}
{\cal H}=J_1\sum_{<i,j>} \vec S_i\cdot \vec S_j
+J_2\sum_{<i,k>} \vec S_i\cdot \vec S_k
+J_3\sum_{<i,l>} \vec S_i\cdot \vec S_l,
\end{equation}
where the bonds corresponding to
interactions $J_1$, $J_2$ and $J_3$ are shown in Fig.~1.
We assume $J_1$ is antiferromagnetic and can henceforth set $J_1$ to unity.
For $J_3=0$ the model reduces to the original Shastry-Sutherland model. 
For $J_2=J_3$, the model has infinitely many conserved quantities.
The total spin on each $J_1$ bond commutes with ${\cal H}$ and
thus each eigenstate of the Hamiltonian
can be characterized by the number and position of the triplets present.
These triplets then form 
(in general a diluted) spin-one Heisenberg
model, with nearest-neighbor interactions on the square-lattice.
It is easy to show that this model has 
three phases at $T=0$ with first order transitions between them.
For large negative $J_2$ the ground state is a fully polarized
ferromagnet, for large positive $J_2$ the ground state is equivalent
to the N\'eel ground state of a spin-one Heisenberg model on the square-lattice,
which is rigorously known to be ordered \cite{spinoner}, 
and whose numerical properties are very well established \cite{spinonen}.
In between the ground state is the Dimerized Shastry-Sutherland
singlet phase. At the Dimer to N\'eel transition the
sublattice magnetization jumps from 0 to about $80$ percent of the
classical value. 

Away from the $J_2=J_3$ line, we use
symmetry arguments, Dimer series expansions \cite{weihong}
together with considerations of the classical
limit to discuss the ground-state
phase diagram and properties of this model.
The model with J$_2$ and J$_3$
interchanged can be mapped into the original one by interchanging
the spins on all J$_1$ bonds. Thus the phase diagram is
symmetric with respect to the $J_2=J_3$ line. We will concentrate
our discussion on the region $J_2\ge J_3$. First, let us
compare energies of various classically ordered phases with the
energy of the Dimer phase to get a first view of various phases
and their boundaries.

\begin{figure}
\epsfxsize=80mm
\centerline{\epsffile{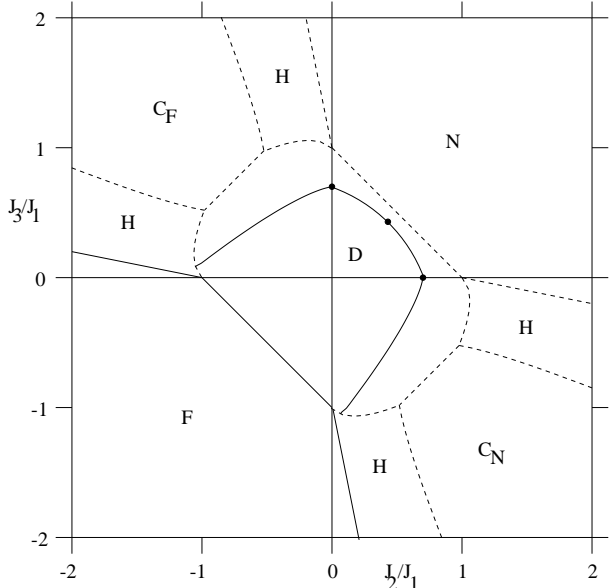}}
\caption[*]{Phase diagram for the model showing ferromagnetic (F),
N\'eel (N), columnar (C$_F$ and C$_N$), helical (H) and dimer (D)
phases. See text for discussion of phase boundaries.}
\label{Fig.2}
\end{figure}

The different phases are shown in Fig.~2, with the Dimer phase denoted
by $D$. 
The magnetically ordered phases are best described with respect
to a rotated lattice where individual spins have four nearest neighbors
(connected by $J_2$ bonds) \cite{shastry}.
This lattice is topologically
equivalent to a square-lattice.
On this lattice, in addition to a ferromagnetic phase (F)
and an antiferromagnetic N\'eel phase (N),
there are columnar (C$_N$ and C$_F$) and helical (H) phases.
The two columnar phases are colinear and equivalent
to each other. In the $C_N$ phase
the spins order antiferromagnetically along one of the axes 
and have period four 
($\uparrow\uparrow\downarrow\downarrow$)
in the perpendicular direction. In the $C_F$ phase, the spins are ferromagnetic
in one direction and have period four in the other direction.
Classically the four helical phases can be mapped onto each other.
In the helical phase for $J_2>-J_3>0$, the helix runs along 
one of the axes, with successive spins rotating by an amount $+\theta$
as one moves along that direction \cite{mila}. Along the perpendicular direction
the change in spin directions alternates between $+\theta$ and $-\theta$.
The angle $\theta$ is non-unique and is one of the
solutions to the equation
\begin{equation}
\cos{(\pi-\theta)}={{2(J_2-J_3)}\over {J_1+\sqrt{J_1-24J_3(J_2-J_3)}}}.
\end{equation}
Defining $x=J_2/J_1$, $y=J_3/J_1$, the columnar, helical, Dimer triple
point in the classical phase diagram is located at $x_{tr}=(45-4\sqrt{6})/36
\approx 0.9778$, $y_{tr}=-(9+4\sqrt{6})/36\approx -0.5222$.
The N\'eel-helical phase boundary is given by $x+5y=1$,
whereas the asymptotic ($x\to\infty$) phase boundary 
between the helical and columnar
phases is given by the equation $y={1\over 19}(-8x+1+O(1/x))$.

Since there are no quantum fluctuations in the ferromagnetic and Dimer
phases and on the boundary between ferromagnetic and helical phases, these
remain true phase boundaries in the model and
are shown by solid lines in Fig.~2. Note that the Dimer-ferromagnetic
boundary is
first order, whereas the Dimer-helical
boundary is second order. The other classical phase
boundaries are shown by dashed lines. They leave an oval-like
Dimer phase in the middle. The first-order Dimer-N\'eel phase boundary can
be determined quite accurately along $J_2=J_3$ to be at $x=0.42957(2)$,
and has been given in previous
numerical studies along $J_3=0$ (and equivalently $J_2=0$) \cite{weihong}.
These points are shown by solid dots and we connect them smoothly
to indicate a first order Dimer-N\'eel phase boundary.

Actually, it is not evident whether the Dimer-N\'eel phase boundary along the
line $J_3=0$ is first or second order, or even whether there is an
intermediate phase between the two.
Albrecht and Mila \cite{mila} have
argued that there is an intermediate helical
phase between the N\'eel and Dimer phases.
Their Schwinger Boson mean-field
treatment leads to the estimate that the N\'eel phase extends only down to
$x\approx 0.91$, whereas the helical phase exists between
$0.61<x<0.91$.
On the other hand, using series expansions,
Weihong et al. \cite{weihong} have argued that 
the N\'eel phase extends down to $x=0.691$ at which point there
is a first order transition to the Dimer phase. 
The finite-size calculations \cite{miyahara,mila}
also do not suggest any helical phases. Though, Albrecht and Mila 
have argued that this is because the helical phases are not properly
accomodated in finite geometries. 

The quantitative
validity of Schwinger Boson calculations is difficult to judge.
One generally expects quantum fluctuations to stabilize colinear
phases. And this could considerably reduce the extent of the helical phases
in the phase diagram. 
In several spin models, where numerical calculations
have been done, the sublattice magnetization of the N\'eel phase 
goes continuously to zero and it is separated from incommensurate
phases by a singlet phase \cite{gsh}.
On general grounds, Ferrer \cite{ferrer} has argued that
the N\'eel phase must be separated from helicoidal phases by
an intermediate spin-liquid phase.
Thus, it is reasonable to assume that along $J_3=0$ there is
a direct transition between Dimer and N\'eel phases.

Along $J_3=0$, Weihong et al. estimate that the Dimer
to N\'eel transition happens at $J_2/J_1=0.691(6)$. Using
d-log Pade approximants to analyze the gap series, we estimate that
it vanishes at $J_2/J_1\approx 0.697(2)$.
Thus, within the uncertainties of the series analysis, this
transition could be continuous. 
Around this value of the couplings, the 
sublattice magnetization series from
the N\'eel side is also consistent with zero \cite{weihong}. 
The primary reason
for believing that the transition is first order is that the energies
for the N\'eel and Dimer phases appear to cross at a non-zero angle. 
However, if the transition is continuous, then very close to the transition,
the N\'eel energy curves should change slope \cite{weihong}.
Thus, given all the numerical evidence, a plausible conclusion is that
the transition is very weakly first order, though it could also be second
order.

Given the above, it is natural to expect
some continuous transitions when $J_3<0$.
To explore this possibility, we have developed
series expansions for the triplet excitations in the dimer
phase to $15$th order for arbitrary $J_3/J_2$ using
the flow equation method \cite{flow}.
For $J_3=0$, the expansion coefficients agree 
with those calculated by Weihong et al.
The unit cell of the lattice contains two dimers, giving rise to
two triplet modes. These modes are almost degenerate throughout the
Brillouin zone and become exactly 
degenerate as $q\to 0$ due to symmetry.
We find that the gap minimum is always at 
$q$ equal to zero even as one moves from the
N\'eel towards the ferromagnetic phase. 
We use d-log Pade approximants to calculate the locus of
points, where the triplet gap closes.
This contour is also depicted by a
solid line in Fig.~2. It marks a boundary at which the Dimer phase
becomes locally unstable, and hence the dimer phase cannot exist
beyond that line. Without studying all eigenstates of our system,
it is not possible to say if some other level crossing transition
leads to a different ground state before we get to this line.
It is plausible to think that at least parts of this line 
represent continuous phase transitions between the Dimer and magnetically 
ordered phases. As seen in Fig.~2, a possible consistent scenario is
that very near the $J_3=0$
line, we have a multicritical point where a second order transition line
meets a first order phase boundary.

These continuous phase transitions between the Dimer phase
and various magnetic phases are rather unusual. They are not
in the conventional $2+1$-dimensional
O(3) universality class as expected for the
non-linear sigma models \cite{qnls}. This is evident from the fact that
in the Dimer phase, the ground state
remains unchanged and hence the correlation length remains of order unity.
In contrast, for a generic dimerized spin system, the correlation
length gradually grows and diverges as the gap goes to zero \cite{sgh}.
The continuous phase transition, here, is 
somewhat analogous to the density driven
generic phase transitions in the Bosonic Mott insulators \cite{mpaf}.

However, there are some important differences.
Unlike the case of Bosonic Mott insulators, the spectrum appears to be
linear at the transition. Along the $J_3=0$ line, we estimate that the
gap vanishes at $x=0.697(2)$, with an apparent exponent $\nu$ of 
$0.45(2)$. Different d-log Pade approximants show remarkable
consistency with each other. Fig.~3 shows the spectrum, in the reduced 
Brillouin zone, at the transition.
Different ways of analyzing
the series all point to a finite spin-wave velocity and a linear
spectrum. These results suggest that this transition belongs to
a new universality class \cite{footnote1}. 

\begin{figure}
\epsfxsize=80mm
\centerline{\epsffile{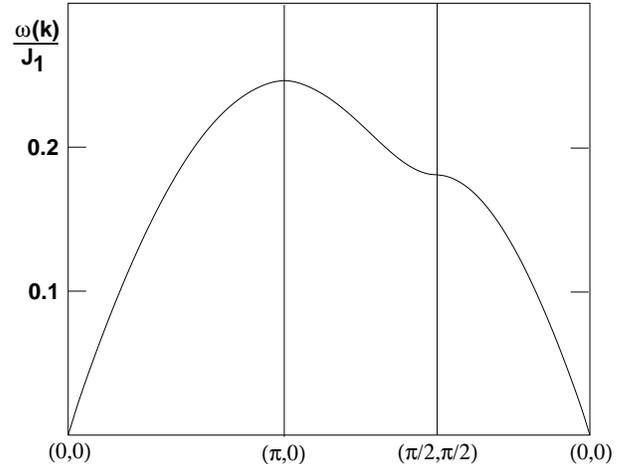}}
\caption[*]{The triplet spectrum along the $J_3=0$ line at $x=0.697$,
where the gap vanishes.}
\label{Fig.3}
\end{figure}

With our present calculations we cannot study the transitions from
the ordered side and thus cannot establish the full nature of these
phase transitions nor can we say anything about the stability of columnar
and helical phases in the overall phase diagram.
Quantum fluctuations can lead to additional singlet (spin-gap) 
phases between the N\'eel and the helical phases and possibly
eliminate helical phases altogether from the N\'eel side
of the phase diagram. 
These N\'eel to singlet phase
transitions should be similar to those found in the $J_1-J_2$
square-lattice Heisenberg model \cite{j1-j2}.

\begin{figure}
\epsfxsize=80mm
\centerline{\epsffile{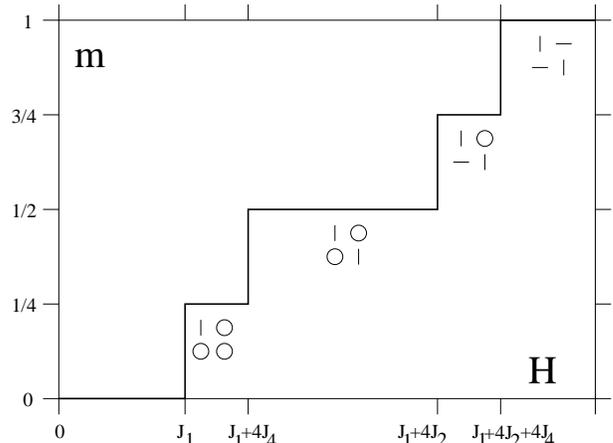}}
\caption[*]{Magnetization as a function of magnetic field. Note that
the plateaus are valid for $J_4<J_2<0.43 J_1$.}
\label{Fig.4}
\end{figure}

When a magnetic field is applied to the Dimer ground state along the
special line J$_2$=J$_3$, the resulting magnetization
is shown in Fig.~4. The triplet excitations,
aligned by the field, have no dispersion,
but a nearest neighbor repulsion. Thus they form a simple Wigner
crystal (or a Bosonic Mott insulator) at one half the saturation
magnetization. 
If we add an additional weak antiferromagnetic coupling between
neighboring horizontal $J_1$ bonds (and similarly neighboring vertical 
$J_1$ bonds)
of the form $J_4(\vec S_1+\vec S_2)\cdot(\vec S_3+\vec S_4)$,
the triplets remain dispersionless but they now have an
additional second neighbor repulsion. This leads to additional
Wigner crystal phases at one and three quarter fillings and hence plateaus
in the magnetization at one and three quarters of the saturation value.
In Fig.~4, we also show the ordering pattern of the Wigner crystal on
different plateaus by showing four $J_1$ bonds. 
A line denotes a ($S^z=1$) triplet on the bond whereas a circle denotes 
a singlet.

As we move away from the $J_2=J_3$ line,
the triplet develops dispersion. It is useful
to think of the problem in terms of the Bose Hubbard model \cite{bose},
with the ($S^z=1$) triplet representing hard core Bosons.
These Bosons have a strong nearest-neighbor repulsion and a
weak diagonal (second-neighbor) and further-range hopping. 
At half-filling, the strong repulsion will clearly lead
to a Wigner crystal and hence the magnetization plateaus
will remain. However, the transition between the
magnetization plateaus may now be second order. 
Indications of such plateaus also exist in the finite size calculations of
Miyahara and Ueda for $J_3=0$ \cite{miyahara}. However, due to the
finite size, all plateaus appear discontinuously in their study. The
question of whether there will be additional magnetization
plateaus at other rational fillings perhaps including valence bond
states, as in one dimension \cite{affleck}, deserves further attention.

We now comment on the materials:
One would naively expect the SrCu$_2$(BO$_3$)$_2$
system to be close to $J_3=0$,
a limit that has been considered by other authors \cite{miyahara,weihong}. 
The ratio of J$_2$ to J$_1$ has been placed in the literature \cite{miyahara}
close to the Dimer to N\'eel transition. Even
though this transition may be first order, it would be very
weakly so, due to the vicinity of the multicritical point.
In this sense, this material may allow us to study a special
quantum critical point, not the generic N\'eel to singlet
quantum phase transition. However, this transition may be 
unstable to the generic $O(3)$ transition, when the special eigenstate
of Shastry and Sutherland is not a true eigenstate due to some small
perturbations. It would be interesting to study this crossover theoretically.
An interesting problem could be the instability
of these special transitions due to spin-lattice couplings.

Magnetization plateaus have been observed in the material at 
one eighth and one quarter of the saturation
magnetization.
The exactly soluble model suggests that these phases may be
regarded as simple Wigner crystals of local triplets.
The question of whether the models
away from $J_2=J_3$ will have magnetization plateaus at other
rational fractions, or whether other couplings
such as $J_4$ are needed for this, deserves further attention. 

We thank Ian Affleck, Leon Balents and Matthew Fisher for discussions.
This work is supported in part by the National Science Foundation
grants PHY94-07194 and DMR96-16574 and by the Deutsche Forschungsgemeinschaft,
SFB 341 and SPP 1073.





\begin{references}
\bibitem{footnote1}
The estimated value of $\nu$ increases
slowly as one moves towards the ferromagntic phase and is about $0.73$
along the negative $J_3$ axis.
\end{references}
\end{document}